\documentclass[aps,prc,superscriptaddress,twocolumn,nofootinbib]{revtex4}
\usepackage{amsmath}
\usepackage{epsfig}
\usepackage{xspace}

\usepackage{xspace}
\usepackage{graphicx}
\usepackage[latin1]{inputenc}
\usepackage{relsize}
\usepackage{axodraw}
\usepackage{epsfig}
\usepackage{amssymb}
\usepackage{mathrsfs}
\usepackage{mathcomp}

\newcommand{\dipsy}{{\smaller DIPSY}\xspace}

\newcommand{\si}[0]{\ensuremath{\sigma}}

\newcommand{\ph}[0]{\ensuremath{\phi}}

\newcommand{\Ps}[0]{\ensuremath{\Psi}}
\newcommand{\ga}[0]{\ensuremath{\gamma}}

\newcommand{\al}[0]{\ensuremath{\alpha}}
\newcommand{\ep}[0]{\ensuremath{\varepsilon}}

\preprint{arXiv:xxxx.yyyy [hep-ph]\\
LU-TP 11-xx\\
MCnet-11-xx\\
\today}                            

\begin{document}
\title{Correlations and Fluctuations in the Initial State of high energy Heavy Ion Collisions}

\author{Christoffer Flensburg}
\affiliation{Dept. of Astronomy and Theoretical Physics, Lund university, Sweden}

\date{\today}
\begin{abstract}
Initial states of high energy heavy ion collisions are studied using a dipole model through the \dipsy event generator that dynamically includes saturation together with the fluctuations and correlations of the BFKL cascade. The eccentricities $\ep_{1,2,3,4}$ are calculated at RHIC and LHC. Predictions are made for correlations and fluctuations in rapidity of the eccentricities, and conventional theoretical approximations are tested. A large set of initial state Au-Au, Cu-Au and Pb-Pb collisions have been generated and are published online.
\end{abstract}
\maketitle
 
\section{Introduction}
\label{sec:intro}
The study of azimuthal correlations and their origin plays a big role in high energy heavy ion collisions. Often the azimuthal correlations are explained by an anisotropy in the transverse plane of the state just after collision, that propagates to the final state through collective effects~\cite{Ollitrault:1992bk}.

Traditionally these studies have been made with average transverse shapes as function of impact parameter $b$, but lately more and more studies have incorporated event-by-event fluctuations of the interaction shape~\cite{Alver:2008zza, Bhalerao:2011bp}. The first sign was that fluctuations are necessary to explain the large elliptic flow in central Cu-Cu collisions seen at PHOBOS~\cite{Alver:2006wh}, but recently also fluctuation driven observables such as triangular flow~\cite{:2008gk, Alver:2009id, Abelev:2008un, Alver:2010gr, Petersen:2010cw, Schenke:2010rr} and directed flow~\cite{Teaney:2010vd,Gardim:2011qn} have been studied. Some of the approaches to model these fluctuations are:

The colour glass condensate model~\cite{Iancu:2003xm}, for example implemented in the KLN model~\cite{Drescher:2006ca}, calculates quantities from an average transverse gluon density, evolved through rapidity. This approach does not include the event-by-event and region-by-region fluctuations and correlations in BFKL.

NeXus~\cite{Drescher:2000ha} is an initial state event generator based on a Gribov-Regge model of hadronic collisions, and is now combined with the hydrodynamical model SPheRIO~\cite{Aguiar:2001ac} to make the full final state event generator NeXSPheRIO~\cite{Takahashi:2009na}. UrQMD, Ultra-relativistic Quantum Molecular Dynamics~\cite{Petersen:2010cw,Petersen:2011fp} uses flux tube excitation and fragmentation to generate initial states. Another model is AMPT~\cite{Lin:2004en, Xu:2011fe}, a multi-phase transport model, using HIJING~\cite{Gyulassy:1994ew} for initial states, where nucleon-nucleon collisions are found from an eikonal formalism.

NeXus, UrQMD and HIJING all use binary collision between the nucleons, and thus do not take saturation in the cascade into account properly.

In this letter the dipole model introduced in~\cite{Avsar:2005iz,Avsar:2006jy,Flensburg:2008ag,Flensburg:2011kk}, with the Monte Carlo implementation \dipsy, will be used to describe the states at time $t=0$. This dipole model is based on BFKL in impact parameter space, and includes all the fluctuations and correlations from a partonic cascade while still taking saturation into account in both evolution and interaction.

The model is briefly introduced in section~\ref{sec:model} and results of common quantities such as eccentricities $\ep_{1,2,3,4}$ follow in section~\ref{sec:results}. In section~\ref{sec:fluct}, the effects of saturation on the BFKL correlations and fluctuations are studied, and section~\ref{sec:conclusions} ends the paper with concluding remarks.

\section{The model}
\label{sec:model}
The model and its Monte Carlo implementation \dipsy are described in detail in~\cite{Flensburg:2011kk}. Below follows a short summary of the key components, and why it is well suited for the analysis of azimuthal eccentricities in high energy heavy ion collisions.

\dipsy is an event generator using colour dipoles in transverse space. The two starting dipole states, representing the colliding particles, evolve in rapidity towards each other to meet at the interaction rapidity $y_0$. The interacting dipoles in the virtual cascade are traced back towards their valence parents, separating the real gluons, referred to as backbone gluons, from virtual fluctuations.

The evolution and interaction amplitudes are based on leading logarithm BFKL, and is corrected for several nonleading effects:
\begin{itemize}
\itemsep 0mm
\item Running $\al_s$. The scale is set by the dipole sizes.
\item Energy conservation. The full 4-momentum and recoils of each parton are tracked, and $p_+$ and $p_-$ ordering is required, simulating non-singular terms in the splitting function and ``energy scale terms''.
\item Confinement is included through a gluon mass, suppressing large dipole emissions.
\item Saturation. Unitarisation provide saturation in the interaction, and the ``dipole swing'' allows for merging of gluon chains in the cascade.
\end{itemize}
The running $\al_s$ and the lightcone momentum ordering makes the model take the essential parts of non-leading logarithmic effects into account, while still including the fluctuations from a BFKL description. Multiple interactions together with the dipole swing gives a description of merging and splitting gluon chains at any rapidity, independently of where the interaction rapidity $y_0$ is set.

A proton in \dipsy starts out as a triangle of dipoles, that then evolves in rapidity to form more complicated states before collision. A heavy ion is here described by a Wood--Saxon distribution of nucleons with a hard core~\cite{DeJager:1987qc,Rybczynski:2010ad}, where each nucleon is described by a triangle just like a proton. This set of triangles in impact parameter space is then evolved as one state giving a web of gluon ladders and loops connecting the nucleons.

The backbone gluons describe the $t=0$ state, which will be used to study properties such as eccentricities in this letter. The gluonic states can also be used as initial condition in a final state evolution model. For this purpose, a large set of collisions of several reactions has been generated and published online at \href{http://home.thep.lu.se/~christof/DIPSYEvents/} {http://home.thep.lu.se/~christof/DIPSYEvents/}.

\section{Results}
\label{sec:results}

\subsection{$\ep_n$ at RHIC and LHC}
\label{sec:epsn}
The calculation of $\ep_n$ in this letter will follow the method in~\cite{Alver:2010gr} with the formula
\begin{equation}
\ep_n = \frac{\sqrt{\langle r^2 \cos (n \phi) \rangle^2 + \langle r^2 \sin (n \phi) \rangle^2}}{\langle r^2 \rangle}. \label{eq:epsr2}
\end{equation}
Here the averages are over all real gluons in a rapidity slice $\eta \in [-1,1]$, and $r$ and $\phi$ for the gluons are determined with respect to the center of gravity.
\begin{figure}[t]
  \includegraphics[width=0.7\linewidth,angle=270]{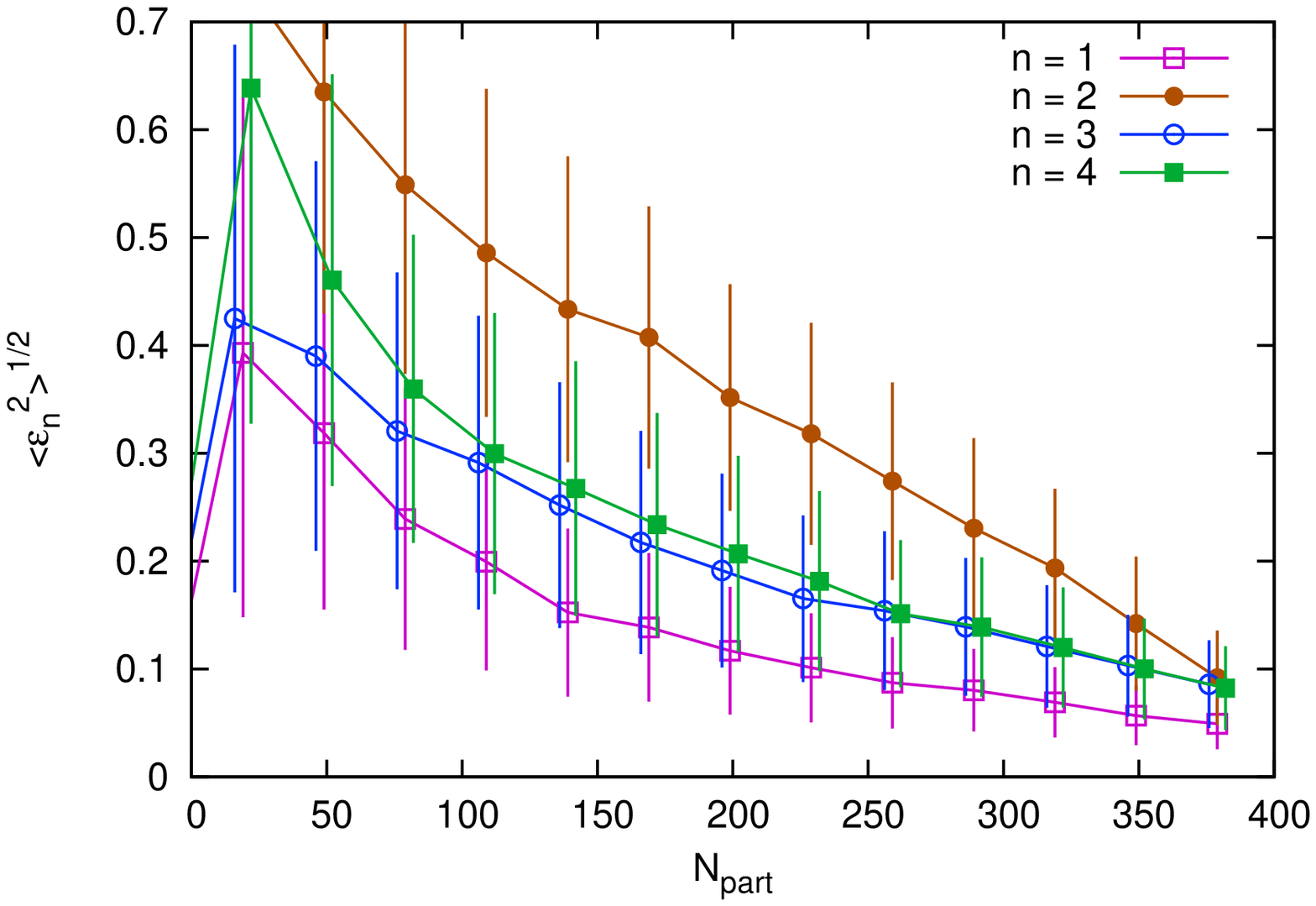}
  \caption{\label{fig:default} (Colour online) The eccentricities at Au-Au collisions for $\sqrt{s_{NN}} = 200$~GeV as function of impact parameter $b$. The error bars show $\si_{\ep_n}= \sqrt{\langle \ep_n^2 \rangle - \langle \ep_n \rangle^2}$.}
\end{figure}
\begin{figure}[t]
  \includegraphics[width=0.7\linewidth,angle=270]{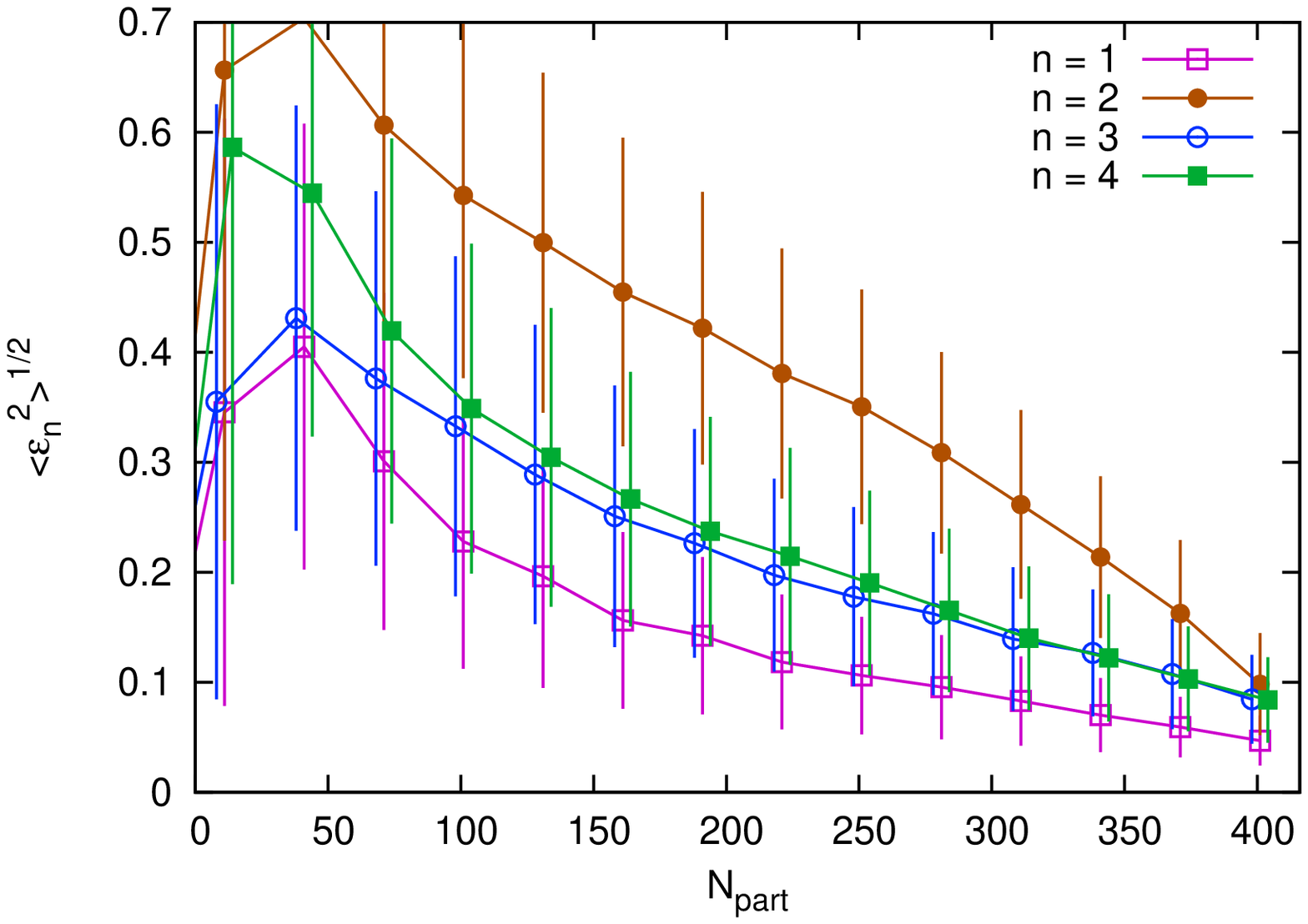}
  \caption{\label{fig:LHC} (Colour online) The eccentricities at Pb-Pb collisions for $\sqrt{s_{NN}} = 2.76$~TeV as function of impact parameter $b$. The error bars show $\si_{\ep_n}= \sqrt{\langle \ep_n^2 \rangle - \langle \ep_n \rangle^2}$.}
\end{figure}

The eccentricities will be shown as function of the number of participants. The spectators can be identified in \dipsy as the nucleons that have no interacting emissions.

Also the formula
\begin{equation}
\ep_n = \frac{\sqrt{\langle r^n \cos (n \phi) \rangle^2 + \langle r^n \sin (n \phi) \rangle^2}}{\langle r^n \rangle} \label{eq:epsrn}
\end{equation}
is in use~\cite{Petersen:2010cw,Alver:2010dn} for the eccentricities. The difference between the two definition is essentially only a constant factor $(n+2)/4$ as argued in~\cite{Alver:2010dn}, and this ratio is confirmed by \dipsy data.

$\ep_1$ however, is calculated according to~\cite{Teaney:2010vd,Gardim:2011qn} from
\begin{equation}
\ep_1 = \frac{\sqrt{\langle r^3 \cos ( \phi) \rangle^2 + \langle r^3 \sin ( \phi) \rangle^2}}{\langle r^3 \rangle} . \label{eq:epsr1}
\end{equation}

For a closer comparison to observables, $\sqrt{\langle \ep_n^2 \rangle}$ will be shown rather than $\langle \ep_n \rangle$~\cite{Alver:2010dn}. This adds about 15\%, 7\%, 13\% and 10\% to $\ep_{1,2,3,4}$ respectively.

The eccentricities for Au-Au collisions at $\sqrt{s_{NN}} = 200$~GeV as function of impact parameter are shown in fig.~\ref{fig:default}, and the corresponding plot for Pb-Pb at $\sqrt{s_{NN}} = 2.76$~TeV is shown in fig.~\ref{fig:LHC}. The error bars show the fluctuations $\si_{\ep_n} = \sqrt{\langle \ep_n^2 \rangle - \langle \ep_n \rangle^2}$ and, assuming not too large fluctuations in the final state evolution, the Au-Au ratio $\si_{\ep_2}/\sqrt{\langle \ep_2^2 \rangle}$ agrees with data for $\si_{v_2}/v_2$ from PHOBOS~\cite{Alver:2007qw} and STAR~\cite{Sorensen:2006nw} for all centralities.
\begin{figure}[ht]
  \includegraphics[width=0.7\linewidth,angle=270]{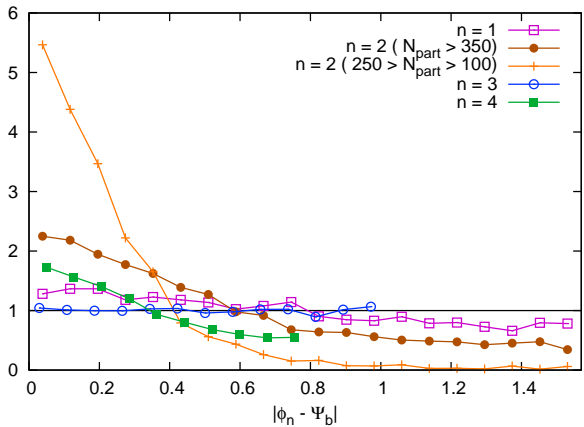}
  \caption{\label{fig:angle} (Colour online) The angle between the event plane and the participant plane for  $N_{part} > 100$ Au-Au collisions at $\sqrt{s_{NN}} = 200$~GeV. The normalisation is such that uncorrelated angles yield a constant 1.}
\end{figure}

The angle $\phi_n$ between the participant plane of the $n$:th moment and the event plane can be seen in fig.~\ref{fig:angle}. As seen in~\cite{Petersen:2010cw}, $\ph_2$ is strongly correlated to the event plane angle $\Ps_b$ at medium impact parameters where the geometric effect is strong, and significantly less correlated at central events where the ellipticity is more fluctuation driven. $\ph_3$ is independent of impact parameter, and $\ph_1$ only weakly correlated.

\section{Correlations and fluctuations}
\label{sec:fluct}
With the dynamics from the saturated BFKL cascade, \dipsy is expected to describe correlations and fluctuations between different rapidity slices within one event.
\begin{figure}[t]
  \includegraphics[angle=270,width=0.49\linewidth]{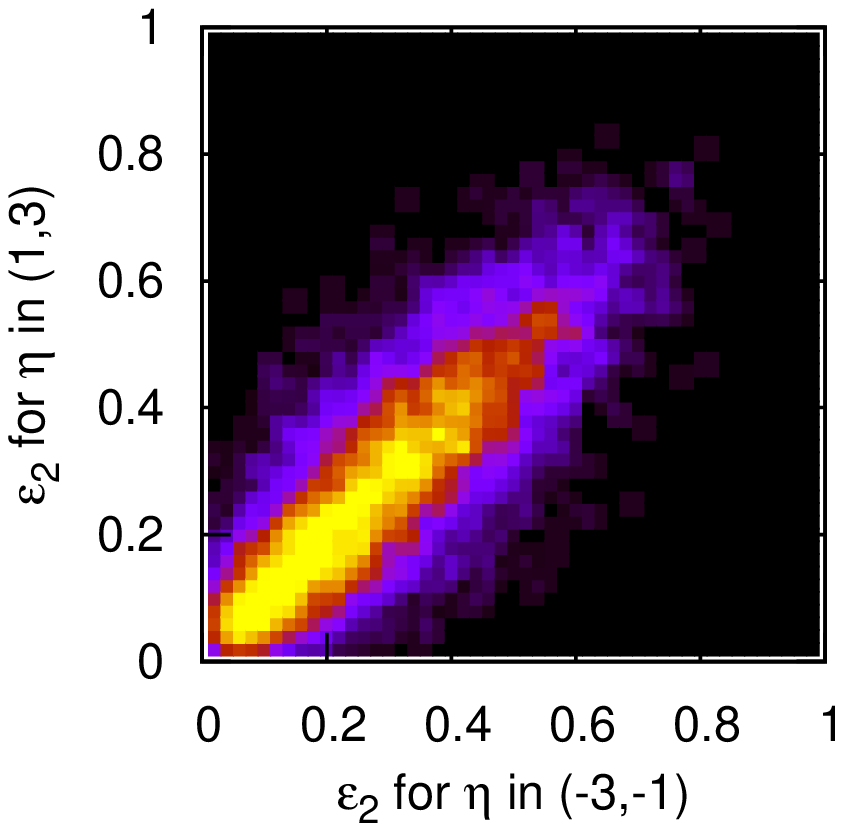}
  \includegraphics[angle=270,width=0.49\linewidth]{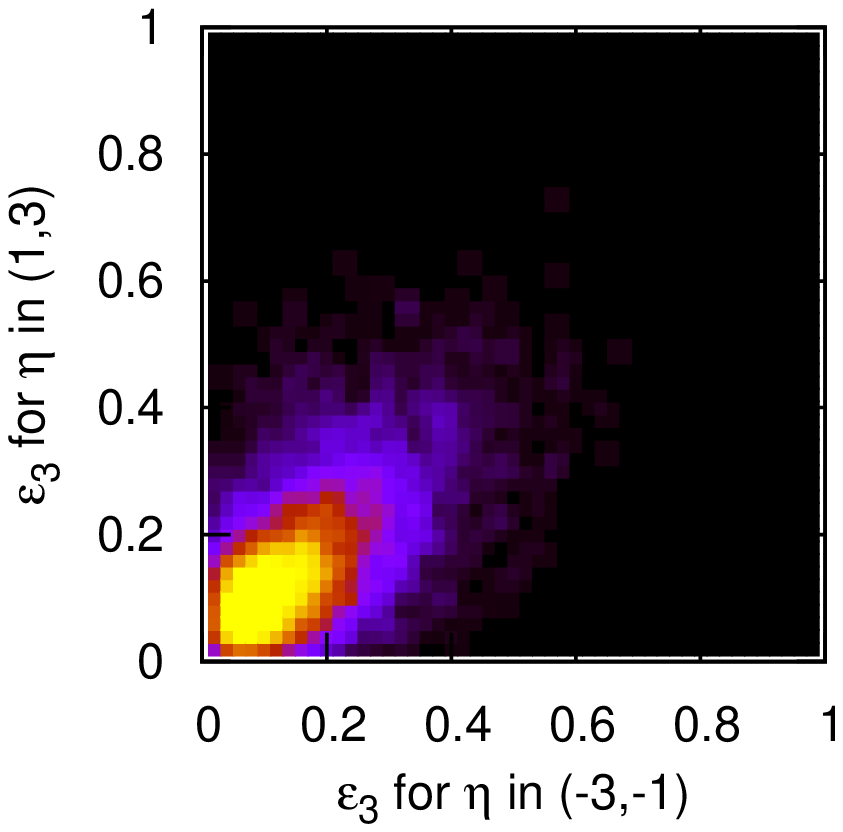}
  \caption {\label{fig:corr} (Colour online) The correlation between $\ep_n$ in the forward region and backward region at $N_{part} > 100$ Au-Au collisions for $\sqrt{s_{NN}} = 200$~GeV. $n=2$ (left) and $n=3$ (right).}
\end{figure}

One such observable is the ellipticity in the forward region as function of the ellipticity in the backward region, seen in figure~\ref{fig:corr}. The ellipticity is seen to be more correlated over rapidity than the triangularity. The correlation between $\ep_n$ in $\eta \in (1,3)$ and in $\eta \in (-3,-1)$ can be quantified with the correlation coefficient
\begin{equation}
\rho_{\ep_n^\text{F},\ep_n^\text{B}} = \frac{\langle \ep_n^\text{F} \ep_n^\text{B} \rangle - \langle \ep_n^\text{F} \rangle \langle \ep_n^\text{B} \rangle}
{\sqrt{\langle (\ep_n^\text{F})^2 \rangle - \langle \ep_n^\text{F} \rangle^2} \sqrt{\langle (\ep_n^\text{B})^2 \rangle - \langle \ep_n^\text{B} \rangle^2}}
\end{equation}
where the index $\text{F}$ and $\text{B}$ implies the eccentricity in the forward, $\eta \in (1,3)$, or backward, $\eta \in (-3,-1)$, region. $\rho_n$ is 1 if the eccentricities are perfectly correlated, and 0 if they are completely uncorrelated. At $N_{part} > 100$ RHIC Au-Au, $\rho_{\ep_n^\text{F},\ep_n^\text{B}}$, or $\rho_n$ for short, is 0.49, 0.81, 0.57, 0.58 for $n = 1,2,3,4$ respectively. Notice that $\ep_2$ is more correlated over rapidity as its origin is a systematic effect. The same is seen at LHC Pb-Pb, where $\rho_{1,2,3,4}$ are 0.62, 0.88, 0.71, 0.69 respectively.

\begin{figure}[t]
  \includegraphics[width=0.7\linewidth,angle=270]{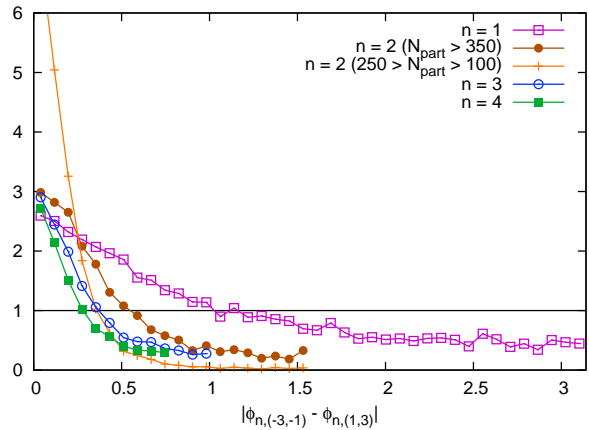}
  \caption {\label{fig:AngleCorr} (Colour online) The angle between $\ph_n$ in the forward region and backward region at $N_{part} > 100$ Au-Au collisions for $\sqrt{s_{NN}} = 200$~GeV. The normalisation is such that uncorrelated angles yield a constant 1.}
\end{figure}

Looking at the angle between the participant plane in the forward and backward region in fig.~\ref{fig:AngleCorr}, it is seen that the strong correlation in the ellipticity is coming from the mid-centralities, where the systemtic effect is strongest.

Experimental verification of these correlations in the angle and magnitude of the corresponding anisotropic flow observables would reinforce the picture of flow coming from systematic effects and fluctuations in the initial transverse geometry.

\begin{figure}[t]
  \includegraphics[width=0.6\linewidth,angle=270] {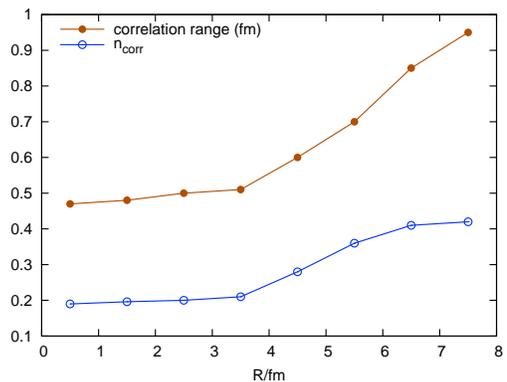}
  \caption{\label{fig:fluxtubes} (Colour online) The number of correlated gluons $n_{corr}$, and the correlated gluons typical transverse range, in pseudorapidity $(2,3)$ due to a trigger gluon in pseudorapidity $(-3,-2)$ in $b=0$ Au-Au with $\sqrt{s_{NN}} = 200$~GeV. $R$ is the distance from the center of the collision.}
\end{figure}

A more direct measure of the positional correlation between rapidities is to measure the transverse density of gluons in the forward rapidity slice triggered by a gluon in the backward rapidity slice. An extra $n_{\text{corr}}$ gluons are seen on top of the background in the corresponding transverse position in the forward slice, due to the gluon chain from the trigger gluon that may pass through also the opposite rapidity bin. This correlation is analogous to a flux tube in other models. Fig.~\ref{fig:fluxtubes} shows central ($b=0$) Au-Au collisions with trigger particles at a distance $R$ from the center of the collision.

It is seen that in the center of the collision, the most dense region, fewer gluons are correlated to the trigger gluon. This is a sign of saturation, as the trigger gluon chain does not propagate unhindered to the other rapidity slice, but has a large probability of merging with other chains before that, giving fewer correlated gluons in the other slice. In the peripheral region of the collisions, $n_{\text{corr}} \approx 0.4$, which is similar to $pp$ collisions in \dipsy. This shows that the approximation of binary collisions, where flux tubes propagate independently through rapidity, breaks down in the dense region of a heavy ion collisions.

The shorter range of the correlation in the center of the collision is again due to the saturated environment preferring smaller dipoles through the dipole swing.



\section{Conclusions}
\label{sec:conclusions}
A new method to generate $t=0$ states of high energy heavy ion collisions has been introduced with \dipsy. The model has been tuned to $pp$ and $\ga^*p$ minimum bias events, and no new parameters have been introduced for heavy ions. \dipsy is based on BFKL and includes all mergings and splittings of gluon chains, describing all fluctuations and correlations in a saturated environment.

$t=0$ events for Au-Au and Cu-Au at RHIC, and Pb-Pb at LHC are generated and avaliable online, see sec.~\ref{sec:model}.

Eccentricities $\ep_{1,2,3,4}$ and their angles $\ph_{1,2,3,4}$ have been studied and give results similar to other models, with fluctuation-driven quantities generally slightly larger. Predictions were made for correlations between eccentricities in the forward and backward regions, which puts the concept of azimuthal flow from initial state geometry to the test. $v_2$ for mid-centrality classes are expected to be significantly more correlated, both in amplitude and orientation, over rapidity than other moments as well as the elliptic flow at central collisions.

Correlations over long range in rapidity are found between the transverse gluon distributions, as is expected from a flux tube approach. Studying head on Au-Au collisions at RHIC energy, the correlation is weaker by a factor 2 in the center compared to the peripherial region of the collision. The weaker correlation is caused by gluon chain mergings, as a gluon chain (or correspondingly a flux tube) passing the trigger rapidity slice, does not neccessarily imply that the same chain (flux tube) pass the other rapidity slice. Also the range of the transverse correlation is shortened a factor 2, due to the smaller average dipole size in a saturated environment, corresponding to a larger saturation scale $Q_s$.

\section{Acknowledgements}
First I wish to thank Gösta Gustafson and Leif Lönnblad, without whom this work could not have been done. The work depends on the model for nucleon positions which Andras Ster helped us include in \dipsy. I am also grateful for valuable discussion with Peter Christiansen, Jamie Nagle and Jean-Yves Ollitrault.

\bibliography{/home/william/people/leif/personal/lib/tex/bib/references,refs}
\end{document}